\documentclass[5p]{elsarticle}
\makeatletter
\def\ps@pprintTitle{%
 \let\@oddhead\@empty
 \let\@evenhead\@empty
 \def\@oddfoot{\centerline{\thepage}}%
 \let\@evenfoot\@oddfoot}
\makeatother

\usepackage{hyperref}
\hypersetup{
  colorlinks,
  citecolor=Violet,
  linkcolor=Red,
  urlcolor=Blue}

\usepackage{longtable}
\usepackage{todonotes}

\usepackage{amsmath}
\usepackage{amssymb}
\usepackage{color}

\begin{document}

\begin{frontmatter}

\title{Graph coarse-graining reveals differences in the module-level structure of functional brain networks}

\author[cs]{Rainer Kujala\corref{cor1}}
\ead{Rainer.Kujala@aalto.fi}
\author[nbe]{Enrico Glerean}
\author[cs]{Raj Kumar Pan}
\author[nbe]{Iiro P. J\"a\"askel\"ainen}
\author[nbe]{Mikko Sams}
\author[cs]{Jari Saram\"aki}
\cortext[cor1]{Corresponding author}

\address[cs]{Department of Computer Science, School of Science, Aalto University, Helsinki, Finland}
\address[nbe]{Department of Neuroscience and Biomedical Engineering, School of Science, Aalto University, Helsinki, Finland}

\begin{abstract}
Network analysis is rapidly becoming a standard tool for studying functional magnetic resonance imaging (fMRI) data.
In this framework, different brain areas are mapped to the nodes of a network, whose links depict functional dependencies between the areas.
The sizes of the areas that the nodes portray vary between studies. Recently, it has been recommended that the original volume elements, voxels, of the imaging experiment should be used as the network nodes to avoid artefacts and biases.
However, this results in a large numbers of nodes and links, and the sheer amount of detail may obscure important network features that are manifested on larger scales. One fruitful approach to detecting such features is to partition networks into modules, \emph{i.e.}~groups of nodes that are densely connected internally but have few connections between them. However, attempting to understand how functional networks differ by simply comparing their individual modular structures can be a daunting task, and results may be hard to interpret. We show that instead of comparing different partitions, it is beneficial to analyze differences in the connectivity between and within the very same modules in networks obtained under different conditions. We develop a network coarse-graining methodology that provides easily interpretable results and  allows assessing the statistical significance of observed differences.
The feasibility of the method is demonstrated by analyzing fMRI data recorded from 13 healthy subjects during rest and movie viewing. While independent partitioning of the networks corresponding to the the two conditions yields few insights on their differences, network coarse-graining allows us to pinpoint \emph{e.g.} the increased number of intra-module links within the visual cortex during movie viewing.
Given the computational and visualization challenges due to increasing resolution and accuracy of brain imaging data, we expect that the importance of methods such as network coarse-graining will become increasingly important in helping to interpret the data.
\end{abstract}


\end{frontmatter}

\textbf{
This is the pre-peer reviewed version of the following article: Kujala, R., Glerean, E., Pan, R. K., J\"a\"askeläinen, I. P., Sams, M., Saram\"aki, J. (2016), Graph coarse-graining reveals differences in the module-level structure of functional brain networks. European Journal of Neuroscience, 44: 2673–2684, which has been published in final form at \url{http://dx.doi.org/10.1111/ejn.13392}. 
This article may be used for non-commercial purposes in accordance with Wiley Terms and Conditions for Self-Archiving.
}

\section{Introduction}

Methods of network science are increasingly used for analyzing functional magnetic resonance imaging (fMRI) data~\cite{Newman2010,Sporns2009,Power2011Neuron,papo2014functional}.
In this framework, fMRI time series are mapped to a network, where the nodes correspond to different brain areas and the links between nodes indicate dependencies between the blood-oxygen-level dependent (BOLD) time series of these areas.
Whenever the dependency between the BOLD signals corresponding to two brain areas is strong enough, they are thought to be functionally related.

Network analysis has revealed many insights on the functional structure of the brain.
For instance, on the scale of network nodes, the use of various network centrality measures has allowed consistent detection of certain hub regions~\cite{vandenHeuvel2013}. At the network level,  the entire structure of functional brain networks has been found to be of the `small-world' type~\cite{Watts1998,Eguiluz2005,Salvador2005}, meaning that the average number of steps required to reach any other node in a network is low while the network is locally clustered.
In addition to revealing such general properties, the network framework has also been used for investigating how brain dynamics depend on different stimuli~\cite{lahnakoski2012naturalistic}, mental health~\cite{Achard2012a,AlexanderBloch2012,Glerean2015}, and age~\cite{Meunier2009Age}.

Despite being widely applied, the network approach to brain functional networks can still be considered as somewhat immature, and methodological variations persist in the literature~\cite{Stanley2013,Garrison2015}. In particular, there is no standard set of brain areas to be used as network nodes; rather, different studies use different definitions of what constitutes a node~\cite{Stanley2013}.
Approaches for defining network nodes range from the use of anatomical atlas-based Region-Of-Interest (ROI) parcellations to the use of the original fMRI imaging voxels of a few mm$^3$ in size~\cite{Stanley2013}.
Typically, the number of nodes in atlas-based definition schemes is of the order of $10^2$, while the number of nodes in voxel-based node definition schemes is of the order of $10^3-10^4$.
These differences between node definitions make comparison of results across studies challenging.

The use of small number of nodes is understandable as it makes network analysis easier, and a low number of nodes even allows
meaningfully visualizing the networks.
However, such low resolution also means that the nodes of a network may cover multiple functionally specific areas.
As the representative BOLD signal for each node is typically computed as the average over the node's spatial extent,
 this can lead to significant loss of information and node-level signals that are not representative of true function.
The recommended remedy is to instead use individual voxels as the network nodes~\cite{Stanley2013}.
Given that with high-field (\emph{e.g.} 7 Tesla) fMRI it has been possible to reach a spatial resolution smaller than 1 mm$^3$
~\cite{yacoub2008high}, the volumes that voxel-based nodes represent can be expected to decrease in the future, while their number increases.

Constructing and analyzing networks that consist of voxel-level nodes is challenging because of the large number of nodes and links.
While computing node-wise centrality measures or global network characteristics may be possible for voxel-level networks,
understanding the \emph{the overall organization of a network's links} becomes difficult. Because the number of potential links is of the order of millions, mere visualization of such networks is extremely challenging, and \emph{e.g.}~comparing two networks on a link-to-link basis becomes meaningless.

One approach for going around this problem is to look at network structure on an intermediate level between the level of nodes and the level of the entire network. This intermediate level is often approached by splitting the network into \emph{modules}, \emph{i.e.}~groups of nodes that are densely connected internally but have few connections between them~\cite{Fortunato2010}.
Modules are typically discovered using stochastic algorithms that partition the network into non-overlapping node groups. For examples of  applications of module detection to functional brain network analysis, see \cite{Meunier2009,Meunier2010,Uehara2012,Power2011Neuron,betzel2016dynamic,sporns2016modular}.

Any observed modular structure of functional brain networks may arise from experimental conditions (different stimuli), or reflect more persistent, underlying features of the subject brain. Some recent studies have investigated how network modules differ between healthy controls and patients suffering from schizophrenia~\cite{AlexanderBloch2012}, patients with autism spectrum disorders~\cite{Glerean2015}, or patients who are comatose~\cite{Achard2012a}.
However, due to various intricacies in the detection of network modules and the difficulty of comparing different module partitions, results have been difficult to interpret and their statistical significance has remained elusive.
Although one can statistically approach the stability of modules~\cite{Moussa2012,Glerean2015} or the similarity of partitions between groups of networks~\cite{AlexanderBloch2012}, verifying the significance of specific differences in modules remains an open problem.
Therefore, there is a need for appropriate methods for analysing and comparing functional network structure at the intermediate level of modules. We argue that instead of focusing on how the network modules themselves differ between groups of networks, it is more fruitful and methodologically sound to assess the differences in the numbers of connections between and within a fixed set of modules that is used as a frame of reference for all groups. The boundaries of these modules are  defined using a specific network or group of networks, and applied as such to the rest of available networks. This coarse-graining approach allows transparent, statistically verifiable investigation of module-level differences between networks.

Below, we first demonstrate the difficulties of comparing network modules with the help of a toy example, and proceed to show how the coarse-graining approach overcomes these problems. Then, to show the applicability of our approach in practice, we apply it to networks constructed from fMRI data recorded for 13 subjects during rest and movie viewing.

\section{Comparing functional networks: coarse-graining and alternative approaches}
\label{sec:coarse-grain-toy}

Perhaps the simplest approach for studying differences in the link structure of functional brain networks is to investigate how the existence or weight of individual links differs between groups of networks.
In this case, assessing the statistical significance of the observed differences is straightforward.
However, with voxel-based functional brain networks that have thousands of nodes and millions of links this approach is impractical for several reasons.
First, the number of statistical tests that need to be performed reaches millions, requiring a large amount of computational resources.
Second, the spatial locations of functionally specific parts of the brain differ across people even after imaging data have been transformed into standard coordinates.
Consequently, there is no one-to-one correspondence between specific nodes and links across subjects.
Third, visualizing all differences between groups of networks becomes impossible. %

Because of the above, some studies have focused on quantifying differences between networks at the level of modules that have been discovered using stochastic network partitioning algorithms~\cite{AlexanderBloch2012,Glerean2015}.
It is worth noting that there is no universally agreed definition for network modules~\cite{Fortunato2010} -- rather, each algorithm introduces its own definition of a network module that dictates how it partitions a network and how modules should look like. Subsequently, different algorithms give different results. Therefore, proper interpretation of module partitions requires a profound understanding of the underlying mathematical ideas and of the actual implementation of the algorithm.

\begin{figure}
\begin{center}
\includegraphics[width=0.95\linewidth]{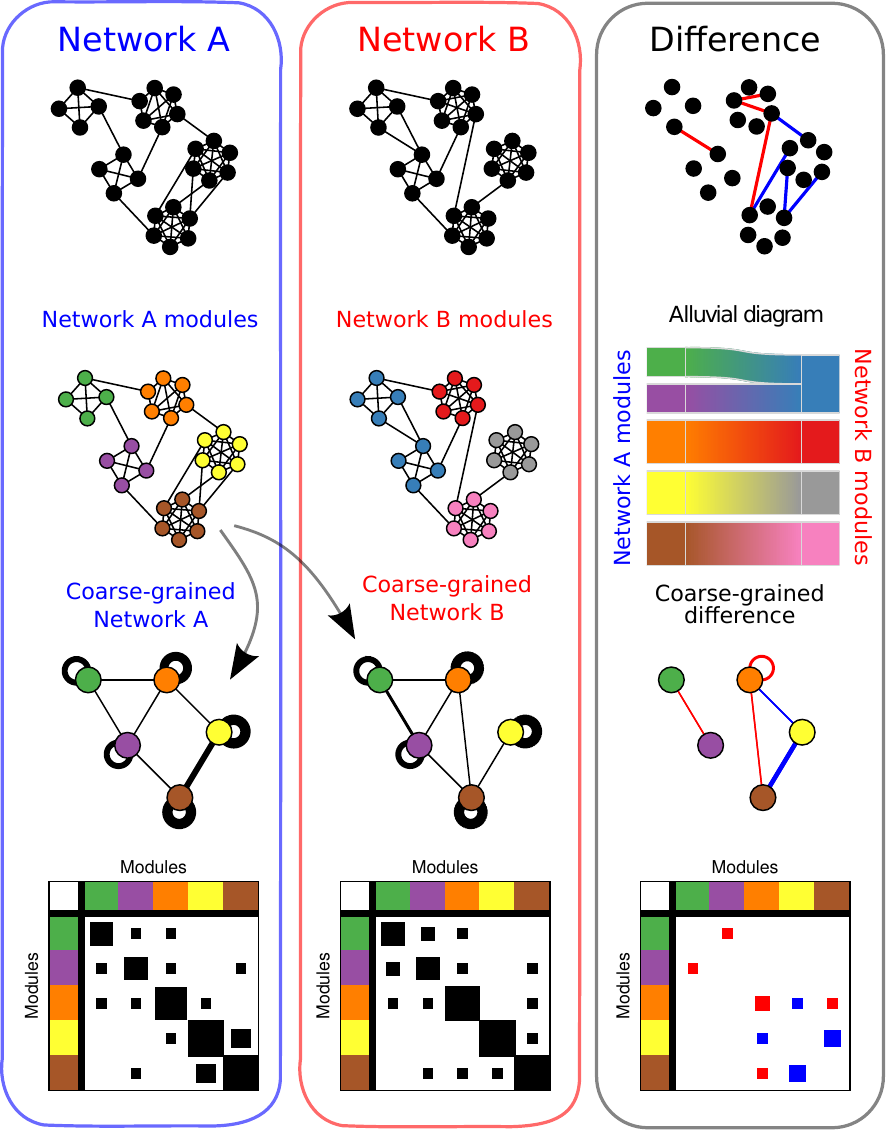}
\end{center}
\caption{\textbf{Network modules, difficulties in comparing them, and the strengths of network coarse-graining.}
\textbf{Top row} shows two networks A (left) and B (middle) that differ from each other only by a relocation of four links (right); blue links are present in A but missing in B, whereas red links are there in B but not in A.
\textbf{Second row} shows the network modules corresponding to networks A and B as identified by the Louvain algorithm that optimizes modularity~\cite{Newman2004, Blondel2008}. On right, the differences in the discovered modular structure are visualized as an alluvial diagram~\cite{Rosvall2010}: the left side of the diagram represents the modules of network A and the right side represents the modules of network B. Ribbons connecting the left and right side show how the modules of A and B match each other in terms of their node composition: it is seen that the only difference between A and B is that the green and violet modules of A correspond to only one module in network B (blue).
This difference arises from the addition of a single link between the green and violet modules.
To the contrary, the three links between the yellow and brown modules in network A that are relocated in B do not give rise to differences in the modules;  the rest of changes (see top row, right) do not produce any changes in modules either.
These examples highlight the difficulty in inferring differences in the link structure of networks based on the modular structure alone.
\textbf{Third row} shows coarse-grained versions of networks A and B and their difference, the coarse-graining being based on the modules detected for A. Here, the width of the link between two modules corresponds to the number of links between their constituent nodes in the original network.
Similarly, the width of the arc around each module represents the number of links within the module. The differences between the coarse-grained networks are shown on the right. For the blue link, there are more between-module links in A than B, and for the red links, B has more between-module links.  Note,how the coarse-grained difference network is able to compactly summarize the differences between networks A and  B.
\textbf{Bottom row} shows the same information as the third row, but in the matrix form. Each (square) element of the matrix corresponds to a module, and the area of the square is proportional to the number of links between modules (off-diagonal) or within modules (diagonal). The row and column colors correspond to the modules of network A.}
\label{fig:intro-example}
\end{figure}

If interpreting what module partitions signify can be difficult, it is even far more difficult to interpret differences between partitions.
Often, small differences in a network's link structure give rise to large differences in the partitioning of the network into modules. Yet, at times,  major structural differences result in little or no differences in the discovered modules.
We illustrate this with Fig.~\ref{fig:intro-example}. In this toy example, we consider two networks, A and B, that have a clearly modular structure. Both networks share the same set of nodes and their link structures are almost the same, with 4 links being different wired differently in B. We then apply the Louvain algorithm~\cite{Blondel2008}, one of the most popular methods for detecting network modules, to the two networks separately. While the removal of three links between  the yellow and brown modules of  Network A results in no differences in the modules detected in network B, the addition of only one extra link between the two smaller green and violet modules in Network A forces them to merge into one large blue module in Network B. This simple example illustrates that it is not straightforward to infer the underlying differences in the overall organization of links by comparing only the network partitions: the cause and the apparent effect size can be disproportionate.

When dealing with experimental, noisy fMRI data the challenges are even greater. The randomness in module partitions that arises from the stochasticity of the algorithms is further amplified by the noise in the data, which obscures real structural differences between networks. In addition, there are no statistical frameworks for directly assessing the statistical significance of differences between partitions -- \emph{e.g.}, is the merging of the two network A modules into the the larger blue network B module statistically significant or just due to chance?
Thus, even when the structure of network modules may shed some light on the functioning of the brain under different conditions, drawing definite conclusions from differences between network partitions is very difficult.

We have seen that directly comparing module partitions is difficult, as is making comparisons at the fine-grained level of individual links. However, as we will show below, combining both points of view yields fruitful results. In particular, we argue that first producing a fixed set of modules, whose boundaries will be the same in all networks, and then comparing inter- and intra-module connections across networks reveals their differences more clearly. Conceptually, this corresponds to network coarse-graining, where each module corresponds to a node of the coarse-grained network, and the number of links between two modules in the original network corresponds to the weight of the link in the coarse-grained network. The number of internal links within each module is taken into account as the weight of the self-link (connection from the node to itself) of each module-node of the coarse-grained network.
The usefulness of the coarse-graining approach is demonstrated in Fig.~\ref{fig:intro-example} with the same toy networks as earlier.
Notably, the comparison of coarse-grained networks reveals  differences in a more transparent way than attempting to compare modules. In addition, when groups of networks are to be compared, the coarse-graining approach allows statistical testing of the differences in the mean number of links within a module or between two modules.

\section{Materials and Methods}
\subsection*{Participants}
The participants were 13 healthy native Finnish speakers (ages 22-43 years, 2 females, 2 left-handed, no neurological or psychiatric history, no hearing impairments, normal vision). The ethical committee of the Hospital district of Helsinki and Uusimaa granted permission for this study which was conducted in accordance with the guidelines of the declaration of Helsinki. Each subject gave written informed consent prior to participation.

\subsection*{Stimulus paradigm}
The stimulus used in this paper has also been used previously~\cite{Lahnakoski2012,Salmi2014} and consisted of an edited
version of the Finnish movie `The Match Factory Girl' (Aki Kaurism\"aki 1990).
The film was projected on a semi-transparent screen behind the subject's head and the audio track was delivered via plastic tubes through porous earplugs.
Each subject went through three sessions with the following order: resting state (15 minutes, 450 volumes), free viewing of the film (22\,min 58s, 689 volumes), resting state (15 minutes, 450 volumes).
In this study we analyze data recorded during rest before viewing the movie, and during movie viewing.
After preprocessing, the movie session was truncated to match the length of the rest session to avoid any biases from different scan durations.%

\subsection*{Data acquisition}
MR imaging was conducted on a 3.0T GE Signa Excite MRI scanner, with a quadrature 8-channel head coil.
A total of 29 functional gradient-echo planar axial slices (thickness 4 mm, 1 mm gap between slices, in-plane resolution 3.4 mm $\times$ 3.4 mm, imaging matrix 64 $\times$ 64, TE 32 ms, TR 2000 ms, flip angle 90$^\circ$,).
T1-weighted images were also acquired (TE 1.9 ms, TR 9 ms, flip angle 15$^\circ$, SPGR pulse sequence) with in-plane resolution of 1 mm $\times$ 1 mm, matrix size 256 $\times$ 256 and slice thickness 1 mm with no gap.

\subsection*{Preprocessing}

Preprocessing of the fMRI data was carried out with FSL (release 4.1.6 www.fmrib.ox.ac.uk/fsl).
The first 10 volumes of each session were discarded from the analysis.
Motion correction was performed with McFlirt and the data were spatially smoothed using a Gaussian kernel with 6\,mm full-width half maximum, and high-pass filtered with a 100\,s cutoff.
Functional data were co-registered with FLIRT to the anatomical image allowing 7 degrees of freedom.
Furthermore, the data were registered from the anatomical space to the MNI152 2mm standard template (Montreal Neurological Institute), allowing 12 degrees of freedom.
The signal was bandpass filtered with a passband of 0.01--0.08\,Hz in accordance with standard functional connectivity procedures.
To control for motion artefacts, motion parameters were regressed out from the data with linear regression (36 Volterra expansion based signals, see Ref.~\cite{Power2014}).
As it is known that head motion affects connectivity results, we controlled for motion with framewise displacement~\cite{Power2012}: all subjects had at least 95\% of time points under the suggested displacement threshold of 0.5\,mm.
For this reason, we decided not to use the scrubbing technique and utilize all time points.
When looking at the individual mean frame-wise displacement, there was no significant difference between conditions ($p$=0.2732).
Finally, to further control for artefacts, voxels at the edge between brain and skull where the signal power was less than 2\% of the individual subject€'™s mean signal power were excluded from the analysis.
This resulted in 5562 6-mm isotropic voxels of brain grey matter covering the whole cerebral cortex, subcortex and cerebellum.
After removing the first and last 15 data points due to bandpass filtering artifacts~\cite{Power2014}, we obtained for each 6-mm voxel a BOLD time-series with 410 time points corresponding to a duration of 13\,min 40\,s.

\subsection*{Network construction}

The functional dependency of two voxels $i$ and $j$ can be measured in many ways, given their BOLD time series $s_i(t)$ and $s_j(t)$~\cite{Smith2011}.
As there is no consensus on the best measure, we opt for simplicity and use the Pearson correlation coefficient, which has been shown to capture a major proportion of pairwise dependencies in fMRI data~\cite{Hlinka2011}.
For each subject and condition, we then compute a correlation matrix $\textbf{R}$, whose elements $r_{ij}$ are the estimated Pearson correlation coefficients between voxels $i$ and $j$.
Given that we have 2 different conditions and 13 subjects, this yields 26 correlation matrices in total.

There are several ways of constructing networks from such matrices by thresholding them so that only chosen elements remain.
One common approach is to use a constant threshold value, so that only nodes pairs whose correlation coefficients (link weights) exceed this value are connected by a link. Another typical approach is to include a fixed fraction of the strongest links in the network.
Here, we adopt the latter approach as it has been shown to provide more stable estimates of various network measures~\cite{Garrison2015}.

We construct networks from correlation matrices as follows: for each matrix, we begin with an empty network, where each node corresponds to a voxel.
We then first compute the maximal spanning tree (MST) of the correlation matrix, and insert the corresponding links to the network. As the MST connects all nodes, this guarantees that no nodes or groups of nodes remain isolated in the network; isolated modules would cause technical difficulties in the later stages of our pipeline.
Next, we sort all correlation coefficients, and insert links corresponding to the strongest positive coefficients until the network contains $\frac{1}{2}N(N-1)\rho$ links  in total, where $\rho \in [0,1]$ is a pre-defined network density.
As the end result, we obtain 26 undirected, unweighted networks that all share the same set of nodes, and have the same number of links.

\subsection*{Selection of network density}

Choosing the fixed density for the thresholded networks is not a straightforward problem. There are no commonly-accepted criteria for
choosing an optimal density. If the density is very low, too much information is discarded and features of interest may remain hidden. On the other hand, the presence of too many links may obscure relevant structures. Sometimes this problem can be overcome by investigating  network structure across different network densities~\cite{AlexanderBloch2012,Lord2012}. However, carrying out detailed structural
analysis of a large number networks of different densities quickly becomes overwhelming. Then, selecting a reasonable specific network density may be a better option. This is also the case in this study, where we compare the module-level structure of groups of networks in detail.

\begin{figure}
  \begin{center}
    \includegraphics[]{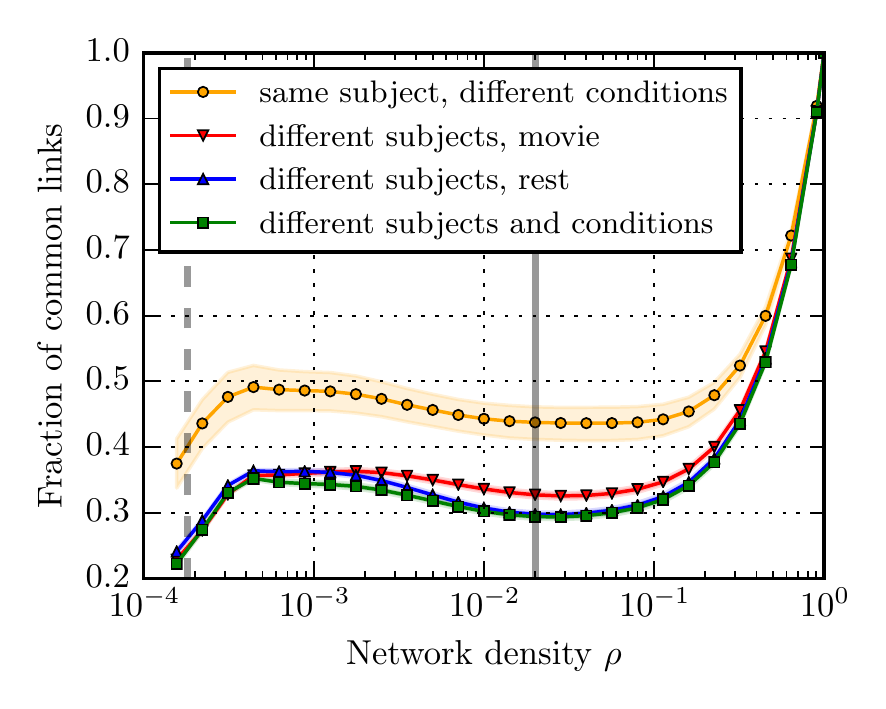}
    \caption{Fraction of common links as a function of the network density, averaged over pairs of networks representing same subject in different conditions $(\circ)$, different subjects viewing the movie($\triangledown$), different subjects at rest ($\triangle$), and different subjects under different conditions ($\square$). Shaded areas denote 95\% bootstrap confidence intervals of the mean.
    The dashed vertical line denotes the network density that corresponds to the inclusion of the minimal spanning tree (MST) (= $\frac{1}{N_\text{nodes}-1}$), and the solid vertical line denotes the 2\% network density which we use in the actual module-level analyses.
    The plot shows that the data is strongly paired: the average fraction of common links for networks representing the same subject is significantly higher than the average fraction of common links for network pairs corresponding to different subjects.
    However, if we consider pairs of networks corresponding to different subjects, networks measured for movie-viewing tend to be more similar than other network pairs.
    }
    \label{fig:frac-com-links}
  \end{center}
\end{figure}

To guide our choice of network density, we investigate the similarity of pairs of networks as a function of the network density.
If two networks share the same set of nodes and have the same number of links, the most straightforward approach to measure their similarity between is to count the number of shared links. This is the approach we adopt.
Given two networks $G$ and $G'$ that both have $\frac{1}{2}N(N-1)\rho$ links, we monitor how the fraction of links $f(G,G')$ common to the networks changes with the network density $\rho$. The fraction of common links $f(G,G')$ is defined as the number of  shared links divided by the total number of links in one network:
\begin{equation}
f(G, G') = \frac{\text{number of common links in $G$ and $G'$}}{\frac{1}{2}N(N-1)\rho}  \in [0,1].
\end{equation}
In Fig.~\ref{fig:frac-com-links} we show  $f$  as a function of the network density when averaged over pairs of networks, so that each pair represents (i) the same subject in different conditions, (ii) different subjects in the rest condition, (iii) different subjects in the movie condition, and (iv) different subjects in different conditions.
For all cases, the fraction of common links first increases until $\rho \approx 0.1\%$; one might envision that at this density, a common ``backbone'' shared by networks is well captured.
Then $f$ decreases until $\rho \approx$ 2\%, after which it begins to monotonously increase as the networks become denser and more and more links are necessarily shared.

Based on the observed behaviour of $\rho$, we pick the value $\rho=2\%$ to be used in all subsequent analyses, as it appears to maximize non-trivial variation between network pairs (of course, there is more variation for excessively low $\rho$).
This value was also used in Ref.~\cite{AlexanderBloch2012} for networks with smaller numbers of nodes.
In our data, the 2\% network density translates to 309\,303 links in each thresholded network; for different networks, this on average corresponds to a correlation coefficient threshold of 0.56$\pm$ 0.04 (std) (no statistically significant difference between conditions).

Fig.~\ref{fig:frac-com-links} also reveals some further insights.
First, the network similarity $f(G,G')$ is remarkably higher for networks corresponding to the same subject in different conditions than for different subjects in the same condition.
This indicates that individual variation dominates over differences caused by different stimuli. Subsequently, it is essential to take the paired nature of the data into account when validating any results statistically.
Second, when the similarity of networks of different subjects is assessed, network pairs corresponding to the movie condition are seen to be more similar than network pairs corresponding to rest condition or different conditions.
This is expected, as the viewing of a well-directed movie stimulus has been found to synchronize the subjects' brains~\cite{hasson2010reliability}, which results in increased functional connectivity compared to the similarities of resting state networks arising from shared functional anatomy and connectivity.

\subsection*{Computation of condition-wise consensus partitions}

\begin{figure}
\begin{center}
\includegraphics[width=\linewidth]{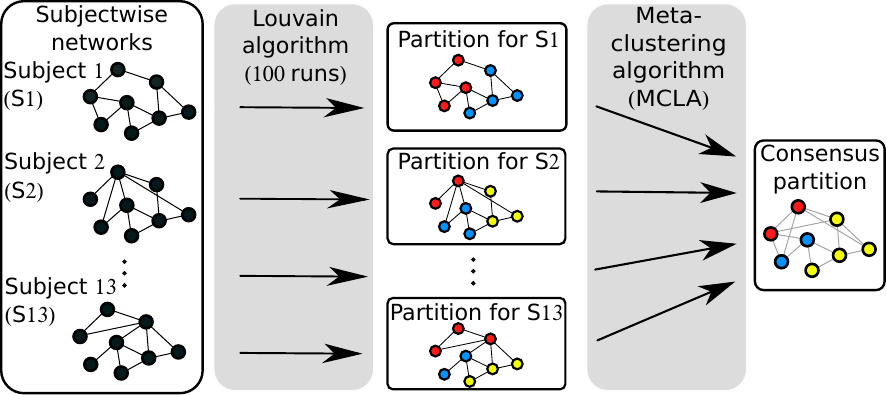}
\end{center}
\caption{Pipeline for computing consensus modules for a networks corresponding to one experimental condition. First, for each subject S1-S13 and their functional network, we run the Louvain algorithm 100 times, and preserve the partition with the best modularity. Then, all 13 best partitions are summarized as a single consensus partition using the MCLA algorithm.}
\label{fig:module-pipeline}
\end{figure}

Given the networks for all subjects S1-S13 and conditions, our next target is to compute representative network partitions for each condition.
As outlined in Fig.~\ref{fig:module-pipeline}, this is realized in two steps:
First, we partition the networks using a popular partitioning algorithm, the Louvain algorithm~\cite{Blondel2008}, which is based on modularity optimization~\cite{Newman2004}.
In more detail, we run the stochastic Louvain algorithm 100 times for each network and select the network partition with the highest value of modularity.
Then, for each condition, we summarize all 13 selected partitions with the help of the MCLA meta-clustering algorithm~\cite{Strehl2003} which yields one representative network partition as an output.
In addition to the network partitions, MCLA requires the user to input a parameter determining the upper bound for the number of modules in the consensus partition.
In this study, the value of this parameter was set to the median number of modules in the 13 partitions.

To summarize, our pipeline transforms the subject-specific networks corresponding to one condition to a single representative \emph{consensus partition} $\mathcal{P}$ consisting of $m$ modules $C_1, ..., C_m$ which in turn are sets of network nodes such that each node belongs to exactly one module.

\subsection*{Comparing groups of coarse-grained networks}
\label{sec:methods-coarse-graining}

The network coarse-graining process briefly introduced in Fig.~\ref{fig:intro-example} is defined as follows:
Given a network $G$ and a partition $\mathcal{P}$ consisting of a set of modules $\{C_1, C_2, ..., C_m\}$, the coarse-graining process yields a matrix $\mathbf{W}$ that has the following properties:
The non-diagonal matrix element $\mathbf{W}_{i,j}$ represents the total number of links between the nodes of $C_i$ and $C_j$. Similarly,
each diagonal element  $\mathbf{W}_{i,i}$ represents the number of links within module $C_i$.

To evaluate differences between experimental conditions, we first coarse-grain each subject's network into its matrix representation $\mathbf{W}^{\text{condition,i}}$, where $i$ stands for the index of the subject.
Then, we average the coarse-grained matrix representations of all 13 subjects over each condition:
\begin{equation}
\langle \mathbf{W} \rangle ^{\text{rest}} = \frac{1}{13}\sum_{i=1}^{13} \mathbf{W}^{\text{rest},i},
\end{equation}
and
\begin{equation}
\langle \mathbf{W} \rangle ^{\text{movie}} = \frac{1}{13}\sum_{i=1}^{13} \mathbf{W}^{\text{movie},i}.
\end{equation}
Then, by investigating the elements of the mean difference matrix $\Delta \mathbf{W} = \langle\mathbf{W} \rangle^{\text{movie}} - \langle \mathbf{W} \rangle ^{\text{rest}}$, we can quantify the level of differences in the numbers of connections between and within modules.
Thus, the value of $\Delta \mathbf{W}_{1,2}$ indicates how many more connections there are on average between modules 1 and 2 in the movie condition than in the rest condition.

To test the statistical significance of our findings, we perform paired permutation tests separately on each of the matrix elements.
As we have a limited number of samples, we use the full permutation distribution yielding in total $2^{13} = 8192$ different permutations.
All $p$-values we report are two-sided, and we correct for multiple comparisons using the original Benjamini-Hochberg (BH) FDR correction~\cite{Benjamini1995}.

\subsection*{Code}
The Python code used in our analysis is freely available at \url{http:github.com/rmkujala/brainnets}.

\section{Results and Discussion}

\subsection*{Consensus modules are similar to previously reported resting-state modules}

The consensus modules computed for the rest and movie-viewing conditions are shown in Fig.~\ref{fig:results}A with different colors on the cortical surface.
A browsable display of each module is available at NeuroVault~\url{http://neurovault.org/collections/1080/}~\cite{gorgolewski2015neurovault}, where the modules are weighted by their consistency as measured by scaled inclusivity~\cite{Steen2011} (see Supporting Information for more details).
In the center of Fig.~\ref{fig:results}A, the matching of the rest and movie consensus partitions is also visualized as an alluvial diagram~\cite{Rosvall2010}.

For the rest condition, we identified 11 consensus modules (Fig.~\ref{fig:results}, left hand side of alluvial diagram, from bottom to top):
1) Limbic (LIM)€" subcortical midbrain structures;
2) Cerebellum/ventro-temporal (CRBL/VT);
3) Default mode (DM);
4) Precuneus (PCUN);
5) Visual (VIS);
6) Auditory (AUD);
7) Salience (SAL);
8) Fronto-Parietal (FP);
9) Visual-extrastriate (VISx);
10) Sensorimotor (SM);
11) Language (LAN).
For the movie condition, we obtained 10 modules:
1) Ventro-temporal limbic (VTL);
2) Cerebellum (CRBL);
3) Default mode (DM);
4) Cuneus (CUN);
5) Visual (VIS);
6) Auditory (AUD);
7) Salience (SAL);
8) Fronto-parietal (FP);
9) Dorsal attention (DA);
10) Sensorimotor (SM).
For details on how the modules were assigned a label, please see Supplementary Information.

There is a good agreement in the literature on the module-level structure of resting networks, which have been identified using various methods such as multidimensional clustering~\cite{Yeo2011}, Infomap graph clustering~\cite{Power2011Neuron}, and independent component analysis~\cite{Smith2009}.
Overall, the resting state consensus modules we obtained are in line with the previously reported resting state modules (see Table S1).

However, there is no general agreement on the module structure during movie viewing, or more generally, during a task.
While one study has found that task and rest are highly similar~\cite{cole2014intrinsic}, another study has found remarkable differences in subcortical, limbic regions as well as primary sensory and motor cortices~\cite{mennes2013extrinsic}. In our case, overall,
the movie consensus modules are similar to the resting state modules, and \emph{e.g.} the dorsal attention module that has been previously reported in resting state studies~\cite{Yeo2011,cole2014intrinsic} is even better identified amongst in our movie consensus partition than in the rest consensus partition (see Tables S1 and S2).
Interestingly, we also identified a VTL subnetwork present only during movie viewing in agreement with~\cite{Glerean2015}, possibly suggesting stronger functional couplings between brain areas involved in the processing of social and emotional events in the movie.

\subsection*{Differences in condition-wise consensus modules are difficult to directly interpret}

As shown in Fig.~\ref{fig:results}A, the consensus network partitions obtained for the movie and rest conditions are broadly speaking similar: for most rest modules, there is a clear counterpart among the movie modules.
At the same time, almost all rest modules overlap with multiple movie modules (and vice versa) -- there are no simple relationships such as one module splitting into two, and the varying amount of overlap between modules results in a diagram that is not straightforward to interpret beyond the clear matches.

As there are no statistical frameworks that can be used for measuring the significance of the relationships between multiple modules, the differences in the alluvial diagram lack statistical validation.
Some insights into the significance of the transitions in the alluvial diagram could be obtained by investigating the consistency of the modules~\cite{Moussa2012, Glerean2015}.
However, these methods do not directly assess the significance of individual splits and joinings of modules between partitions.
Thus, it remains challenging to draw conclusions about network differences based on the alluvial diagram alone.

\subsection*{Network coarse-graining allows simple, statistically verifiable interpretations}

\begin{figure*}
\begin{center}
\includegraphics[width=0.95\textwidth]{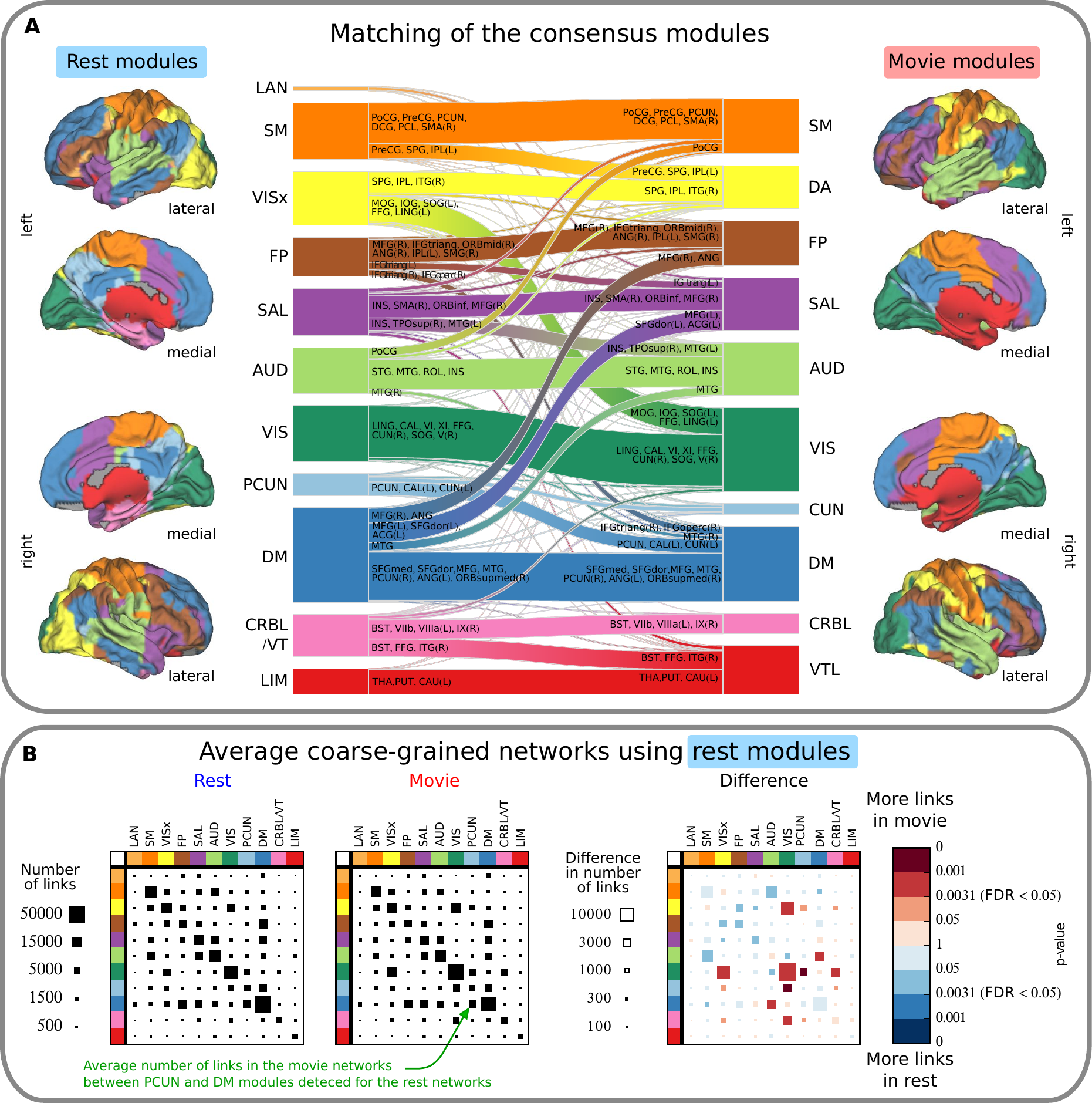}
\end{center}
\caption{
\textbf{Panel A}: Consensus modules and their match between the rest and movie-viewing conditions.
On the left and right, we show the consensus modules obtained for all networks corresponding to the rest and movie conditions, respectively.
The alluvial diagram displayed between the modules of different conditions shows how they relate to one another in terms of node membership.
The height of each module and of the ribbon connecting two modules is proportional to the associated number of nodes.
The colors have been chosen by maximal match between rest and movie modules, to allow visual comparison. The abbreviation of the
module's label is shown next to it (see text for details). Further, each ribbon has also been labeled based the anatomical brain areas that the nodes constituting the ribbon belong to.
For details on the labeling of the modules and ribbons, see Supporting Information.
\textbf{Panel B}: Average coarse-grained networks obtained using the consensus \emph{rest} modules. On the left, the average coarse-grained networks for both rest and movie conditions are shown.
The size of the black rectangles indicate the number of links between modules (within modules for diagonal rectangles). Note that the matrices are symmetric by construction, and thus the upper and lower triangles of the matrices contain the same information.
On the right, the average coarse-grained difference matrix is shown.
Note that the scale for square sizes is different from the left.
The color of each matrix element in the difference matrix indicates the (uncorrected) $p$-value obtained from a mean difference permutation test.
The size and color of a square thus indicate both the associated effect size and the statistical significance of the difference.
The color bar shows the 0.05 FDR threshold computed with the BH procedure.
In total, there are five differences that survive the FDR correction; each of those indicates more connections in the movie condition between certain modules.
}
\label{fig:results}
\end{figure*}

In Fig~\ref{fig:results}B, we show the matrix representations of the average coarse-grained networks corresponding to both conditions, as well as the coarse-grained difference network, where the resting-state consensus modules have been used as the basis for coarse-graining.
As expected, the matrices representing the coarse-grained networks have high values on their diagonals, indicating that the density of links within consensus modules is higher than between them. Off-diagonal element provide an overview of how strongly the modules are connected.

In the coarse-grained difference matrix, we observe multiple elements that survive the 0.05 Benjamini-Hochberg FDR correction. These are also listed in Table~\ref{tab:rest-fdr}.
All surviving elements are positive, indicating that there are more connections between the modules in the movie networks. In particular, the VIS  module displays significantly more external as well as internal connections in the movie condition.
The number of connections between the AUD and DM modules is also increased in the movie condition.
The coarse-graining method thus succeeds in highlighting task-driven changes at visual areas as well as inferior temporal structures.
For similar coarse-graining results where the movie modules are used as the frame of reference, please see Supporting Information.

\begin{table}
\caption{Differences in link numbers within and between rest consensus modules that survive the 0.05 Benjamini-Hochberg FDR correction.}
\label{tab:rest-fdr}
\begin{center}
\begin{tabular}{c c r r}
Module pair $i,j$ & $p$-value & $\Delta W_{i,j}$ & relative increase \\ \hline \hline
VIS,VIS  & 0.00122 & 15184.7 & 46\%  \\  %
VIS,PCUN & 0.00024 & 2624.6 & 60\%  \\
VIS,VISx & 0.00171 & 7865.6 & 47\%  \\
VIS,CRBL/VT & 0.00146 & 3312.1 & 170\% \\
AUD,DM & 0.00293 & 3646.3 & 100\%
\end{tabular}
\end{center}
\end{table}

There are of course some similarities between the coarse-grained difference network and the alluvial diagram presented in Fig.~\ref{fig:results}.
As an example, in both it is seen that the resting-condition VIS module and part of the resting-condition VISx module merge to form the larger VIS module in the movie condition.
A simple explanation for this would be that there are more connections between the VIS and VISx rest modules in the movie condition.
However, as motivated in Sec.~\ref{sec:coarse-grain-toy}, this is not the only possible reason, and further, the statistical significance of the observation cannot be deduced using the module-matching/alluvial-diagram approach.
However, the coarse-grained difference matrix clearly indicates that there is a statistically significant increase (47\%) in the number of links between the VIS and VISx modules in the movie condition.
Thus, while the alluvial diagram can be used for formulating hypotheses on changes in network structure, the coarse-graining process allows verifying that the observed differences are not due to random chance.

In addition to assessing differences in the connectivity between modules, the coarse-graining approach also allows investigating the internal connectivity of modules. This cannot be directly done by comparison of matched modules; further, if the nodes are from the beginning defined as larger entities (\emph{e.g.}~using anatomical atlases), this information is lost.

\subsection*{Methodological considerations}

In this work, we have applied the introduced coarse-graining approach only to the analysis of undirected and unweighted networks.
However, the approach itself is more general and could be easily extended to the analysis of weighted networks that also take into account  the strength of correlation between voxels, or directed networks that depict causal relationships between brain areas.
The coarse-graining approach could also be applied in clinical settings by comparing the module-level differences between different groups of individuals, extending the previous attempts of comparing brain modules~\cite{AlexanderBloch2012, Glerean2015}.

There are many choices to be made in deciding on the particulars of the module detection pipeline, especially the choice of the module detection algorithm (and its parameters, if any). Naturally, these choices will also affect the exact outcome.
However, we expect that regardless of the exact way of defining modules, the benefits of using coarse-graining over comparing partitions
remain. This is because the problems in interpreting the differences between partitions are universal, \emph{e.g.} for any module detection method, there is nearly always a borderline case where the existence or absence of one single link affects boundaries between modules.
Naturally, the choice of a module detection method also affects the resulting coarse-grained networks; however, when comparing across conditions and using the modules of one condition as the frame of reference, the differences should still be straightforward to interpret. Also, which of the conditions is used as the basis for coarse-graining should not be a critical choice: e.g.~ the observed differences in the network structure remain relatively similar for rest and movie modules as basis (see Supporting Information).

It is also essential to point out that, most likely, there is no perfect partition of the brain into distinct areas.
Thus, even though different module detection algorithms yield different network partitions, these may still capture at least some meaningful aspect of the network's organization.
Instead of considering this multiplicity of possible partitions as a problem, it is rather an opportunity, as coarse-graining networks using different partitions can actually turn out to be very useful.
In the same way as one can deduce the shape of a 3D object from its 2D projections, a network's structure can be better understood by investigating its different coarse-grained representations.
Therefore, coarse-graining could be useful even using an anatomical brain atlas as the basis, as the atlas provides a frame of reference that is well known to researchers in the field.

\section{Conclusions}

We developed a coarse-graining method to analyze differences in the modular structure of functional brain networks during rest and task.
The coarse-graining approach focuses on the differences in the connectivity between and within larger brain areas without sacrificing the spatial accuracy of fMRI data already at the network construction stage.
The method yields results that summarizes differences in connectivity on the module level, using a set of modules as a frame of reference across groups or conditions.
In contrary to some alternative approaches studying differences in the module-level connectivity of the human brain, the results produced by our method are both easy to interpret and verify statistically -- they allow to ``see the forest for the trees''.
Because data on the structural and functional human connectome are becoming more and more detailed, we believe such methods will play an increasingly important role in understanding the module-level structure of functional networks.

\section{Acknowledgments}

Computing resources provided by the Aalto Science IT project are acknowledged.
JS acknowledges financial support from the Academy of Finland, project n:o 260427.
EG acknowledges aivoAALTO Project Grant from the Aalto University, doctoral program ``Brain \& Mind".
We thank Onerva Korhonen for comments on the manuscript.

\bibliographystyle{elsarticle-num}
\section*{References}
\bibliography{references}

\clearpage

\onecolumn

\section*{Supporting information}

\renewcommand\thefigure{S\arabic{figure}}
\setcounter{figure}{0}
\renewcommand\thetable{S\arabic{table}}
\setcounter{table}{0}

\subsection*{Text S1: Labeling of the consensus modules}

For labeling the movie and rest consensus modules, we first computed their spatial overlap with known major modules reported in the literature \cite{Power2011Neuron} and \cite{Yeo2011}.
The final module labels were chosen manually and, where possible, matched with the spatial overlap results presented in Tables~\ref{table:labels-rest} and \ref{table:labels-movie}.

In detail, the computation of the spatial overlaps was done as follows:
For each network node, we computed a `stability' measure to describe how well they on average belonged to the consensus module in one condition.
The node-wise stability we used was \emph{scaled inclusivity}~\cite{Steen2011} that has also been previously applied to brain network analyses~\cite{Moussa2012}.
For the movie and rest conditions, we then separately computed the average scaled inclusivity values for each node defined as
\begin{equation}
SI_i = \frac{1}{13} \sum_{J \in \{1,..., 13\}} \frac{|C_i^{\text{REF}} \cap C_i^{\text{J}}| }{ |C_i^{\text{REF}}|} \frac{|C_i^{\text{REF}} \cap C_i^{\text{J}}| }{  |C_i^{\text{J}}| } \in [0,1],
\end{equation}
where $|C_i^{\text{REF}} \cap C_i^{\text{J}}| $ corresponds to the number of nodes that belong both to node $i$'s consensus cluster and node $i$'s cluster in the best partition found for subject $J$.
($|C|$ denotes the number of nodes in cluster $C$)

Our spatial maps of the consensus modules were weighted by the scaled inclusivity values that were separately computed for each condition.
Then we computed the spatial overlap defined as the Pearson correlation coefficient between the spatial maps as done in Refs.~\cite{Smith2009,Glerean2015}.
The code used for computing spatial correlations between brain modules is available at \url{https://github.com/eglerean/hfASDmodules/compare_modules}.

\begin{table}[h!]
\caption{Spatial correlation of the rest consensus modules with reported modules in the literature.}
\label{table:labels-rest}
\vspace{0.4cm}
\begin{tabular}{l |  p{2.5cm} c | p{2.5cm} c | p{2.5cm} c}
Module ID & Yeo et al 2011 & corr. &  Power et al 2011 & corr. & Name given & Abbreviation \\ \hline \hline
rest 1 & Limbic & 0.066 &  Subcortical & 0.84 &  Limbic  &LIM \\
rest 2 & Limbic & 0.21 &   Default mode & 0.1 &  Cerebellum / ventro temporal  & CRBL/VT \\
rest 3 & Default mode & 0.54 &  Default mode & 0.66 & Default mode  &DM \\
rest 4 & Visual & 0.064 &  Memory retrieval & 0.29 &  Precuneus &PCUN \\
rest 5 & Visual & 0.5 &  Visual & 0.6 &  Visual  &VIS \\
rest 6 & Somatomotor & 0.28 &  Auditory & 0.53 & Auditory  &AUD \\
rest 7 & Ventral Attention & 0.27 &  Cingulo-opercular task control & 0.46 & Salience  &SAL \\
rest 8 & Frontoparietal & 0.41 &   Fronto-parietal task control & 0.68 & Fronto-parietal &FP \\
rest 9 & Visual & 0.31 &   Visual & 0.53 & Visual-extrastriate &VISx \\
rest 10 &  Somatomotor & 0.39 &  Sensory/ somatomotor hand & 0.78 & Sensorimotor  &SM \\
rest 11 &  Default mode& 0.079 &   Fronto-parietal task control & 0.13 & Language  &LAN \\
\end{tabular}
\end{table}

\begin{table}
\caption {Spatial correlation of the movie consensus modules with reported modules in the literature.} \label{tab:title}
\label{table:labels-movie}
\vspace{0.4cm}
\begin{tabular}{l | p{2.5cm} c | p{2.5cm} c | p{2.5cm} c}
Module ID & Yeo et al 2011 & corr. &  Power et al 2011 & corr. & Name given & Abbreviation \\ \hline \hline
movie 1 & Limbic & 0.24 &    Subcortical & 0.83 &   Ventro-temporal limbic  & VTL \\
movie 2 & - & - & Default mode & 0.02 &  Cerebellum  & CRBL \\
movie 3 & Default & 0.39 &   Default mode & 0.46 &  Default mode  & DM \\
movie 4 & Visual & 0.14 &    Visual & 0.21 &  Cuneus  & CUN \\
movie 5 & Visual & 0.62 &    Visual & 0.80 &   Visual  & VIS \\
movie 6 & Somatomotor & 0.19 &   Auditory & 0.38 &  Auditory  & AUD \\
movie 7 & Frontoparietal & 0.16 &    Salience & 0.43 &  Salience  & SAL \\
movie 8 & Frontoparietal & 0.39 &    Fronto-parietal task control & 0.59 &  Fronto-parietal & FP \\
movie 9 & Dorsal attention & 0.4 &   Dorsal attention & 0.42 &  Dorsal attention  & DA \\
movie 10 & Somatomotor & 0.42 &   Sensory/ somatomotor hand & 0.71 &  Sensorimotor  & SM
\end{tabular}
\end{table}

\clearpage

\subsection*{Text S2: Node labels}

In the alluvial diagram (Fig.\,4A in the main text), most ribbons connecting the two partitions have labels attached to them.
These labels correspond to the labels of individual nodes, and a label is shown when there are at least 15 nodes with the same label in a ribbon.
The labels for individual nodes were labelled automatically by matching each node with its corresponding automatic atlas labeling (AAL) or Harvard Oxford (HO) label. AAL was used for cerebral cortical areas and HO for subcortical and cerebellar areas.
Furthermore, if both L(eft) and R(ight) parts are present, then no L/R is shown.

\begin{longtable}{l | p{3.5cm} | p{2.5cm} |  p{2.5cm}}
\caption{Abbreviations of node labels} \\
\label{tab:abbreviations} \\
Anatomical name & AAL (cerebral cortex) \& HO (sub-cortex and cerebellum) labels & Major region label & Acronym (as per doi:10.1371/ journal.pone.0005226) \\ \hline \hline
Precentral gyrus (L) & Precentral\_L & Frontal & PreCG(L) \\
Precentral gyrus (R) & Precentral\_R & Frontal & PreCG(R) \\
Superior frontal gyrus (L) & Frontal\_Sup\_L & Frontal & SFGdor(L) \\
Superior frontal gyrus (R) & Frontal\_Sup\_R & Frontal & SFGdor(R) \\
Orbital superior frontal gyrus (L) & Frontal\_Sup\_Orb\_L & Frontal & ORBsup(L) \\
Orbital superior frontal gyrus (R) & Frontal\_Sup\_Orb\_R & Frontal & ORBsup(R) \\
Middle frontal gyrus (L) & Frontal\_Mid\_L & Frontal & MFG(L) \\
Middle frontal gyrus (R) & Frontal\_Mid\_R & Frontal & MFG(R) \\
Orbital middle frontal gyrus (L) & Frontal\_Mid\_Orb\_L & Frontal & ORBmid(L) \\
Orbital middle frontal gyrus (R) & Frontal\_Mid\_Orb\_R & Frontal & ORBmid(R) \\
Opercular inferior frontal gyrus (L) & Frontal\_Inf\_Oper\_L & Frontal & IFGoperc(L) \\
Opercular inferior frontal gyrus (R) & Frontal\_Inf\_Oper\_R & Frontal & IFGoperc(R) \\
Triangular inferior frontal gyrus (L) & Frontal\_Inf\_Tri\_L & Frontal & IFGtriang(L) \\
Triangular inferior frontal gyrus (R) & Frontal\_Inf\_Tri\_R & Frontal & IFGtriang(R) \\
Orbital inferior frontal gyrus (L) & Frontal\_Inf\_Orb\_L & Frontal & ORBinf(L) \\
Orbital inferior frontal gyrus (R) & Frontal\_Inf\_Orb\_R & Frontal & ORBinf(R) \\
Rolandic operculum (L) & Rolandic\_Oper\_L & Frontal & ROL(L) \\
Rolandic operculum (R) & Rolandic\_Oper\_R & Frontal & ROL(R) \\
Supplementary motor area (L) & Supp\_Motor\_Area\_L & Frontal & SMA(L) \\
Supplementary motor area (R) & Supp\_Motor\_Area\_R & Frontal & SMA(R) \\
Olfactory cortex (L) & Olfactory\_L & Frontal & OLF(L) \\
Olfactory cortex (R) & Olfactory\_R & Frontal & OLF(R) \\
Medial superior frontal gyrus (L) & Frontal\_Sup\_Medial\_L & Frontal & SFGmed(L) \\
Medial superior frontal gyrus (R) & Frontal\_Sup\_Medial\_R & Frontal & SFGmed(R) \\
Orbital medial frontal gyrus (L) & Frontal\_Mid\_Orb\_L & Frontal & ORBsupmed(L) \\
Orbital medial frontal gyrus (R) & Frontal\_Mid\_Orb\_R & Frontal & ORBsupmed(R) \\
Gyrus rectus (L) & Rectus\_L & Frontal & REC(L) \\
Gyrus rectus (R) & Rectus\_R & Frontal & REC(R) \\
Insula (L) & Insula\_L & Insula & INS(L) \\
Insula (R) & Insula\_R & Insula & INS(R) \\
Anterior cingulum (L) & Cingulum\_Ant\_L & Cingulate & ACG(L) \\
Anterior cingulum (R) & Cingulum\_Ant\_R & Cingulate & ACG(R) \\
Middle cingulum (L) & Cingulum\_Mid\_L & Cingulate & DCG(L) \\
Middle cingulum (R) & Cingulum\_Mid\_R & Cingulate & DCG(R) \\
Posterior cingulum (L) & Cingulum\_Post\_L & Cingulate & PCG(L) \\
Posterior cingulum (R) & Cingulum\_Post\_R & Cingulate & PCG(R) \\
Parahippocampal gyrus (L) & ParaHippocampal\_L & Occipital & PHG(L) \\
Parahippocampal gyrus (R) & ParaHippocampal\_R & Occipital & PHG(R) \\
Calcarine gyrus (L) & Calcarine\_L & Occipital & CAL(L) \\
Calcarine gyrus (R) & Calcarine\_R & Occipital & CAL(R) \\
Cuneus (L) & Cuneus\_L & Occipital & CUN(L) \\
Cuneus (R) & Cuneus\_R & Occipital & CUN(R) \\
Lingual gyrus (L) & Lingual\_L & Occipital & LING(L) \\
Lingual gyrus (R) & Lingual\_R & Occipital & LING(R) \\
Superior occipital gyrus (L) & Occipital\_Sup\_L & Occipital & SOG(L) \\
Superior occipital gyrus (R) & Occipital\_Sup\_R & Occipital & SOG(R) \\
Middle occipital gyrus (L) & Occipital\_Mid\_L & Occipital & MOG(L) \\
Middle occipital gyrus (R) & Occipital\_Mid\_R & Occipital & MOG(R) \\
Inferior occipital gyrus (L) & Occipital\_Inf\_L & Occipital & IOG(L) \\
Inferior occipital gyrus (R) & Occipital\_Inf\_R & Occipital & IOG(R) \\
Fusiform gyrus (L) & Fusiform\_L & Occipital & FFG(L) \\
Fusiform gyrus (R) & Fusiform\_R & Occipital & FFG(R) \\
Postcentral gyrus (L) & Postcentral\_L & Parietal & PoCG(L) \\
Postcentral gyrus (R) & Postcentral\_R & Parietal & PoCG(R) \\
Superior parietal lobule (L) & Parietal\_Sup\_L & Parietal & SPG(L) \\
Superior parietal lobule (R) & Parietal\_Sup\_R & Parietal & SPG(R) \\
Inferior parietal lobule (L) & Parietal\_Inf\_L & Parietal & IPL(L) \\
Inferior parietal lobule (R) & Parietal\_Inf\_R & Parietal & IPL(R) \\
Supramarginal gyrus (L) & SupraMarginal\_L & Parietal & SMG(L) \\
Supramarginal gyrus (R) & SupraMarginal\_R & Parietal & SMG(R) \\
Angular gyrus (L) & Angular\_L & Parietal & ANG(L) \\
Angular gyrus (R) & Angular\_R & Parietal & ANG(R) \\
Precuneus (L) & Precuneus\_L & Parietal & PCUN(L) \\
Precuneus (R) & Precuneus\_R & Parietal & PCUN(R) \\
Paracentral lobule (L) & Paracentral\_Lobule\_L & Parietal & PCL(L) \\
Paracentra lobule (R) & Paracentral\_Lobule\_R & Parietal & PCL(R) \\
Heschl gyrus (L) & Heschl\_L & Temporal & HES(L) \\
Heschl gyrus (R) & Heschl\_R & Temporal & HES(R) \\
Superior temporal gyrus (L) & Temporal\_Sup\_L & Temporal & STG(L) \\
Superior temporal gyrus (R) & Temporal\_Sup\_R & Temporal & STG(R) \\
Temporal pole (superior) (L) & Temporal\_Pole\_Sup\_L & Temporal & TPOsup(L) \\
Temporal pole (superior) (R) & Temporal\_Pole\_Sup\_R & Temporal & TPOsup(R) \\
Middle temporal gyrus (L) & Temporal\_Mid\_L & Temporal & MTG(L) \\
Middle temporal gyrus (R) & Temporal\_Mid\_R & Temporal & MTG(R) \\
Temporal pole (middle) (L) & Temporal\_Pole\_Mid\_L & Temporal & TPOmid(L) \\
Temporal pole (middle) (R) & Temporal\_Pole\_Mid\_R & Temporal & TPOmid(R) \\
Inferior temporal gyrus (L) & Temporal\_Inf\_L & Temporal & ITG(L) \\
Inferior temporal gyrus (R) & Temporal\_Inf\_R & Temporal & ITG(R) \\
Thalamus (L) & Left\_Thalamus & Subcortex & THA(L) \\
Caudate (L) & Left\_Caudate & Subcortex & CAU(L) \\
Putamen (L) & Left\_Putamen & Subcortex & PUT(L) \\
Pallidum (L) & Left\_Pallidum & Subcortex & PAL(L) \\
Brainstem  & Brain-Stem & Subcortex & BST \\
Hippocampus (L) & Left\_Hippocampus & Subcortex & HIP(L) \\
Amygdala (L) & Left\_Amygdala & Subcortex & AMYG(L) \\
Nucleus accumbens (L) & Left\_Accumbens & Subcortex & NAcc(L) \\
Thalamus (R) & Right\_Thalamus & Subcortex & THA(R) \\
Caudate (R) & Right\_Caudate & Subcortex & CAU(R) \\
Putamen (R) & Right\_Putamen & Subcortex & PUT(R) \\
Pallidum (R) & Right\_Pallidum & Subcortex & PAL(R) \\
Hippocampus (R) & Right\_Hippocampus & Subcortex & HIP(R) \\
Amygdala (R) & Right\_Amygdala & Subcortex & AMYG(R) \\
Nucleus accumbens (R) & Right\_Accumbens & Subcortex & NAcc(R) \\
Cerebellar lobule I-IV (L) & Left\_I-IV & Cerebellum & I-IV(L) \\
Cerebellar lobule I-IV (R) & Right\_I-IV & Cerebellum & I-IV(R) \\
Cerebellar lobule V (L) & Left\_V & Cerebellum & V(L) \\
Cerebellar lobule V (R) & Right\_V & Cerebellum & V(R) \\
Cerebellar lobule VI (L) & Left\_VI & Cerebellum & VI(L) \\
Cerebellar vermis VI  & Vermis\_VI & Cerebellum & VI-vermis \\
Cerebellar lobule VI (R) & Right\_VI & Cerebellum & VI(R) \\
Cerebellar crus I (L) & Left\_Crus\_I & Cerebellum & XI(L) \\
Cerebellar crus I (R) & Right\_Crus\_I & Cerebellum & XI(R) \\
Cerebellar crus II (L) & Left\_Crus\_II & Cerebellum & XII(L) \\
Cerebellar vermis crus II  & Vermis\_Crus\_II & Cerebellum & XII-vermis \\
Cerebellar crus II (R) & Right\_Crus\_II & Cerebellum & XII(R) \\
Cerebellar lobule VIIb (L) & Left\_VIIb & Cerebellum & VIIb(L) \\
Cerebellar lobule VIIb (R) & Right\_VIIb & Cerebellum & VIIb(R) \\
Cerebellar lobule VIIIa (L) & Left\_VIIIa & Cerebellum & VIIIa(L) \\
Cerebellar vermis VIIIa  & Vermis\_VIIIa & Cerebellum & VIIIa-vermis \\
Cerebellar lobule VIIIa (R) & Right\_VIIIa & Cerebellum & VIIIa(R) \\
Cerebellar lobule VIIIb (L) & Left\_VIIIb & Cerebellum & VIIIb(L) \\
Cerebellar vermis VIIIb  & Vermis\_VIIIb & Cerebellum & VIIIb-vermis \\
Cerebellar lobule VIIIb (R) & Right\_VIIIb & Cerebellum & VIIIb(R) \\
Cerebellar lobule IX (L) & Left\_IX & Cerebellum & IX(L) \\
Cerebellar vermis IX  & Vermis\_IX & Cerebellum & IX-vermis \\
Cerebellar lobule IX (R) & Right\_IX & Cerebellum & IX(R) \\
Cerebellar lobule X (L) & Left\_X & Cerebellum & X(L) \\
Cerebellar vermis X  & Vermis\_X & Cerebellum & X-vermis \\
Cerebellar lobule X (R) & Right\_X & Cerebellum & X(R) \\
\end{longtable}

\clearpage

\subsection*{Text S3: Average coarse-grained movie networks}

\begin{figure*}[h!]
\includegraphics[width=1.0\textwidth]{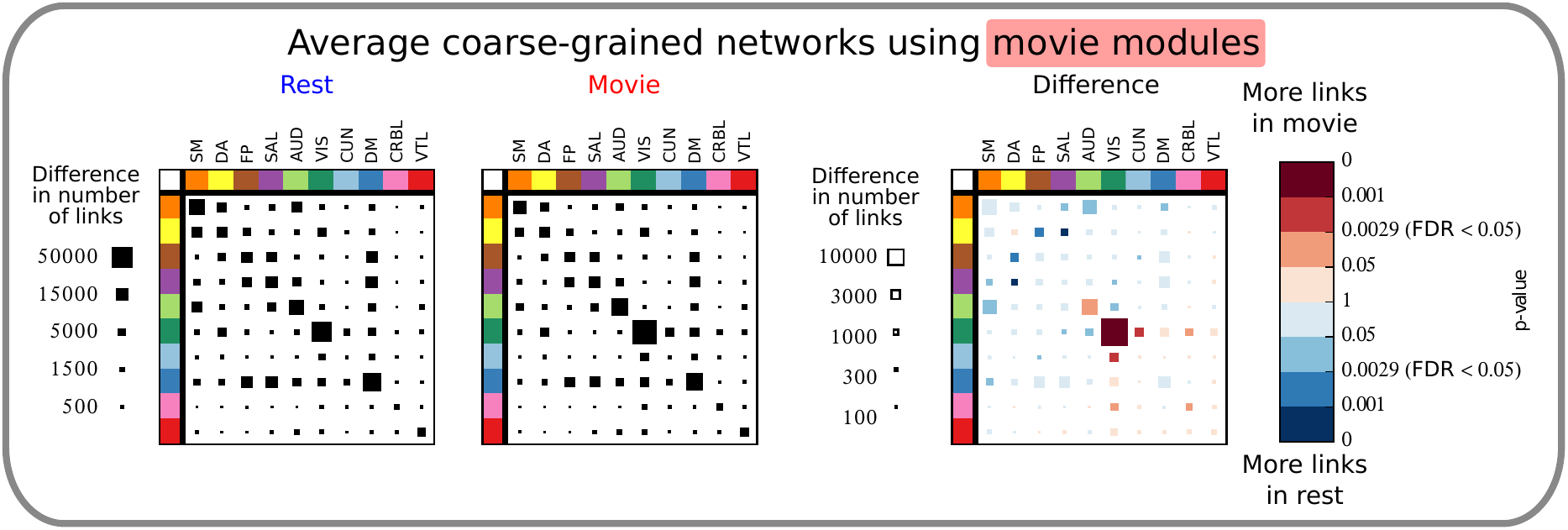}
\caption{
Average coarse-grained networks when the consensus \emph{movie} modules are used in the network coarse-graining process.
On the left, the average coarse-grained networks for both rest and movie conditions are shown.
The size of the black rectangles indicates the number of links between (and within) modules.
Note that the matrices are symmetric by construction, and thus the upper and lower triangles of the matrices contain the same information.
On the right, the average coarse-grained difference matrix is shown.
Note that the scale for the difference matrix has been adjusted for clarity.
The color of each matrix element in the coarse-grained difference indicates the $p$-value obtained from a mean difference permutation test.
The size and color of a square thus indicate both the effect size and the statistical significance of the difference.
In the color bar,  the threshold corresponding to FDR of 0.05 computed with the BH-procedure is also shown.
In total, there are five differences that survive the FDR correction, and in each of those cases there are more connections in the movie condition.}
\label{fig:movie-coarse-grained}
\end{figure*}

Fig.~\ref{fig:movie-coarse-grained} shows the results of the coarse-graining process, when the movie consensus modules have been used as the reference modules for the coarse-graining process.
The differences that survive the 0.05 Benjamini-Hochberg FDR correction are additionally listed in Table~\ref{tab:movie-fdr}.

\begin{table}[h!]
\caption{Differences in link numbers within and between movie consensus modules that survive the 0.05 Benjamini-Hochberg FDR correction.}
\label{tab:movie-fdr}
\begin{center}
\begin{tabular}{c c r r}
Module pair $i,j$ & $p$-value & $\Delta W_{i,j}$ & relative increase (+) / decrease (-) \\ \hline \hline
VIS,VIS  & 0.0010 & 26204.4 & +56\%  \\  %
VIS,CUN & 0.0029 & 2228.9 & +59\%  \\
DA,SAL & 0.0007 & -1120.0 & -54\%  \\
DA,FP & 0.0017 & -1981.3 & -35\% \\
\end{tabular}
\end{center}
\end{table}

\end{document}